\documentclass[aps,prb]{revtex4-1}
\usepackage[T2A,OT1]{fontenc}
\usepackage[english]{babel}
\usepackage{latexsym}
\usepackage{amsbsy}
\usepackage{xcolor}[2007/01/21]
\usepackage[dvipdfm,bookmarks=true,bookmarksnumbered,bookmarksopen]{hyperref}
\hypersetup{pdfpagemode=FullScreen,colorlinks=true,
citecolor=green,unicode,bookmarksopenlevel=3,
pdftoolbar=true,pdfauthor={Helen Gomonay},pdftitle={AFM with
noise}}
\begin{document}
\title{Peculiarities of the stochastic motion in antiferromagnetic nanoparticles}
\author{Helen V. Gomonay}
\affiliation {
 National Technical University of Ukraine ``KPI''\\ ave Peremogy, 37, 03056, Kyiv,
Ukraine}

\author{Vadim M. Loktev}
\affiliation {Bogolyubov Institute for Theoretical Physics NAS of
Ukraine,\\ Metrologichna str. 14-b, 03680, Kyiv, Ukraine}

\begin{abstract}
Antiferromagnetic (AFM) materials are widely used in spintronic
devices as passive elements (for stabilization of ferromangetic
layers) and as active elements (for information coding). In both
cases switching between the different AFM states depends in a
great extent from the environmental noise. In the present paper we
derive the stochastic Langevin equations for an AFM vector and
corresponding Fokker-Planck equation for distribution function in
the phase space of generalised coordinate and momentum. Thermal
noise is modeled by a random delta-correlated magnetic field that
interacts with the dynamic magnetisation of AFM particle. We
analyse in details a particular case of the collinear compensated
AFM in the presence of spin-polarised current. The energy
distribution function for normal modes in the vicinity of two
equilibrium states (static and stationary) in sub- and
super-critical regimes is found. It is shown that the
noise-induced dynamics of AFM vector has pecuilarities compared to
that of magnetisation vector in ferromagnets.
\end{abstract} 
\maketitle
\section{Introduction}
\label{intro} Magnetic nanoparticles are the main constituents of
the nowadays devices for information technology. While the
deterministic dynamics of magnetisation vectors is used for
information coding, the noise-induced stochastic behaviour
facilitates the switching processes and thus is used to increase
the speed of information processing (see, e.g.
\cite{bandiera:202507,Prejbeanu:0953-8984-19-16-165218}). Noise
measurement is a powerful and informative tool for study of the
spintronic effects in different systems
\cite{Aliev:2009PhRvB..79m4423H,Gusakova2011}. The details of the
stochastic processes are also important for the development of
high-quality spin-torque oscillators \cite{krivorotov:2004} and
micropower generators
\cite{Slavin:2007ApPhL..91s2506T,Prokopenko2011ApPhL..99c2507P}.

Theoretical approach to the description of thermal noise in small
\emph{ferromagnetic} (FM) particles was developed in the seminal
papers of W.F.~Brown
\cite{Brown:PhysRev.130.1677,Brown:IEEE15_1196_1979} where the
thermal bath was modeled by fluctuating magnetic field and
corresponding Langevin equations were obtained as generalisation
of the dynamic Landau-Lifshitz-Gilbert equations. This approach
was then extended to the systems in the vicinity of the Curie
point \cite{Garanin:1997PhRvB..55.3050G}, space-inhomogeneous
magnetic vortices \cite{Gaididei:1999PhRvB..59.7010G}, systems
with the coloured noise
\cite{Chubykalo-Fesenko:2009PhRvL.102e7203A,Miyazaki:1998} and FMs
in the presence of spin-polarised current
\cite{Zhang:PhysRevB.69.134416,Apalkov:PhysRevB.72.180405}.

On the other hand, stochastic behaviour of
\emph{antiferromagnetic} (AFM) nanoparticles which are also widely
used in electronic and spintronic devices is studied to a much
lesser extent. The main reason is in the seeming magnetic
neutrality of AFMs which manifests itself in the vanishingly small
or zero net magnetisation, quadratic (in contrast to linear for
FMs) dependence of the internal energy \emph{vs} external magnetic
field etc. Thus, up to now the problem of magnetic relaxation and
thermal noise in AFM particles was in fact reduced to the
description of the effective FM particle with small but nonzero
magnetisation which inevitably appears due to imperfections,
strong external magnetic fields, surface effects etc
\cite{Ouari:PhysRevB.83.064406,Raikher:2008E,Ouari:PhysRevB.81.024412,Mishra:springerlink:10.1140/epjb/e2010-00293-0}.
However, in many cases a peculiar feature of AFM, namely, the
presence of strong exchange coupling between the differently
aligned (mainly in opposite directions) magnetic moments, gives
rise to new dynamic effects that could not be reduced to the
motion of the above mentioned ``stray'' magnetisation. For
example,  in contrast to magnetisation of FM particle, an AFM
(N\'{e}el) vector (defined as the difference between sublattice
magnetisations) can be set into motion not only by external
magnetic field $\mathbf{H}$ but also by its time derivative
$\dot{\mathbf{H}}$ (or curled electric field) \cite{Ivanov:2009}.
Typical frequencies of AFM oscillations are fall into terahertz
range (compared to 1-10 GHz for FM) due to effects of the exchange
enhancement. AFM systems, like FMs, are sensitive to spin-torques
transferred from the spin-polarised current
\cite{Urazhdin:2007,Tsoi-2008,Tang:PhysRevB.81.052401} but the
current-induced AFM dynamics differs significantly from the
current-induced dynamics of FMs \cite{gomo:2010}. Thus, with the
account of perspective to use AFM nanoparticles as alternative to
FM active elements of spin-valves (see,
e.g.\cite{Park:2010arXiv1011.3188P}), theoretical study of thermal
noise in these systems is of importance and of great interest.

In the present paper we generalise the dynamic equations for AFM
nanoparticle to the stochastic Langevin equations that describe
``Brownian motion'' of AFM vector in the presence of thermal
noise. Like in the W.~F.~Brown approach we model the noise as a
fluctuating delta-correlated magnetic field which interacts with
the \emph{dynamic} (induced by the motion of AFM vector
\cite{Andreev:1980,Bar-june:1979E}) magnetisation of AFM. Using
the standard Langevin dynamics technique for multiplicative noise
\cite{Coffey:1998} we derive the Fokker-Planck equations for the
distribution function in the phase space of AFM vector and
corresponding generalised momentum and discuss the peculiarities
of AFM system compared to FM one. As a representative example we
analyse the stochastic behaviour of AFM particle in the presence
of spin-polarised current and find the energy distribution
function in subcritical and supercritical regimes. For the sake of
simplicity we restrict ourselves with the case of a collinear
compensated AFM with two oppositely directed magnetic sublattices.
However, the developed approach allows generalisation for the
multisublatteral AFM, weak ferromagnets etc.

\section{Langevin equations for antiferromagnetic particle}
\label{sec:1} In what follows we consider the fine (nanosized)
magnetic particles whose state can be described with only few
macroscopic vectors\footnote{~Thus, we neglect space
inhomogeneuities of the magnetic subsystem. On the other hand, the
spatial size of the particle is large enough compared with the
correlation length of the magnetic ordering.} $\mathbf{M}_j$:
magnetisation vector ($j=1$) in the case of FM nanoparticle and
vectors of two equivalent sublattice magnetizations ($j=1,2$) in
the case of a collinear AFM.

Let us first discuss the stochastic Landau-Lifshits-Gilbert
equation for the FMs which describes the dynamics of magnetisation
$\mathbf{M}$ subjected to a spin transfer torque
$\mathbf{T}_{\mathrm{STT}}$ at a finite temperature $T$
\cite{Zhang:PhysRevB.69.134416}:
\begin{equation}\label{eq-Landau_Lifshits}
    \dot{\mathbf{M}}=\gamma\mathbf{M}\times\left(\mathbf{H}_{\mathrm{eff}}+\mathbf{h}\right)+\frac{\gamma\alpha_G}{M}\mathbf{M}\times\left[\mathbf{M}\times(\mathbf{H}_{\mathrm{eff}}+\mathbf{h})\right]+\mathbf{T}_{\mathrm{STT}},
\end{equation}
where $\gamma$ is the gyromagnetic ratio, $\mathbf{H}_{\mathrm{eff}}\equiv-\partial w_\mathrm{an}(\mathbf{M})/\partial \mathbf{M}$ is the
 effective (combination of internal and external) magnetic field, $w_\mathrm{an}(\mathbf{M})$ is the magnetic anisotropy energy with Zeeman contribution,
 $\alpha_G$ is the dimensionless damping (Gilbert) constant, the sign $\times$ means
cross-product. The fluctuating magnetic field $\mathbf{h}(t)$ with
a Gaussian stochastic process has the following standard
space-time statistical properties:
\begin{equation}\label{eq_noise_model}
    \langle \mathbf{h}(t)\rangle=0,\quad \langle h_j(t_1)h_k(t_2)\rangle=2D\delta_{jk}\delta(t_1-t_2),
\end{equation}
where $D$ represents the strength of thermal fluctuations whose value is defined from the fluctuation-dissipation theorem. Symbol $\langle \ldots\rangle$ denotes an average taken over all realizations of fluctuating field.

The first r.h.s. term in Eq.~(\ref{eq-Landau_Lifshits}) describes
nondissipative dynamics of magnetisation which is analogue of the
precession of gyroscope with the ``angular momentum''
$\mathbf{M}$. The second r.h.s. term describes dissipation
processes that have the same origin as fluctuating field
$\mathbf{h}(t)$. Both nondissipative and dissipative terms are
subjected to noise, however, the nondissipative noise is additive
while the dissipative one is multiplicative
\cite{Chubykalo:PhysRevB.85.014433}. Delta-correlation of
fluctuating field (\ref{eq_noise_model}) means that the noise
correlation time is much lesser than the characteristic time of
magnetisation response.

The dynamics and kinetics of FM magnetisation with account of
normalisation condition $|\mathbf{M}|=M_0$ (imposed far below the
Curie point) is described with two independent variables that
define space orientation of vector $\mathbf{M}$, in other words,
by the variables of \emph{configuration} space. Full
``mechanical'' energy of FM particle,
$E_{\mathrm{FM}}=w_\mathrm{an}(\mathbf{M})$, also depends upon
orientation of $\mathbf{M}$ and thus can be treated as consisting
of the potential energy only, in contrast to the energy of AFM
particle (see below).

Using the analogy with FMs, one can derive the stochastic equation
for AFMs from the corresponding dynamic equations assuming that
the thermal noise also has the magnetic nature and can be modeled
with the same random field $\mathbf{h}(t)$ (\ref{eq_noise_model}).
This field may originate from fluctuations of \emph{i}) the
surface noncompensated magnetisation for the small particles;
\emph{ii}) magnetisation of the nearest FM layer in spin-valves
and multilayers; \emph{iii}) current that produces additional
magnetic field.

However, the deterministic dynamics in AFMs substantially differs
from that of FM magnetisation and looks like an inertial motion of
a point mass in a potential well. Formally this effect was
demonstrated for AFMs with strong exchange coupling between the
magnetic sublattices \cite{Bar-june:1979E,Andreev:1980}. In this
case the dynamics of a collinear AFM is described by a single AFM
(so called N\'{e}el) vector
$\mathbf{L}\equiv\mathbf{M}_1-\mathbf{M}_2$ of a fixed length
($|\mathbf{L}|=2M_0$).

Deterministic equations of motion for $\mathbf{L}$ could be
obtained either from the set of Landau-Lifshits-Gilbert equations
(\ref{eq-Landau_Lifshits}) for each of magnetic sublattices or,
equivalently, in the framework of Lagrange formalism  (see
\cite{Turov:2001E} for details). The last approach is more
convenient for general analysis, so, we start from the Lagrange
function for AFM particle in the following form
\begin{equation}
\label{Lagrangian_AFM} \mathcal{L}_{\mathrm AFM} =
\frac{m_L}{2}\dot{\mathbf{L}}^2+\gamma
m_L\left[\dot{\mathbf{L}}\cdot(\mathbf{L}\times\mathbf{H})\right]-
w_{\mathrm{an}}(\mathbf{L})+\frac{\gamma^2m_L}{2}(\mathbf{L}\times{\mathbf{H}})^2,
\end{equation}
where $\mathbf{H}$ is an external magnetic field and
$w_{\mathrm{an}}(\mathbf{L})$ is the  energy of magnetic
anisotropy that forms a potential well for AFM vector, $\gamma$,
as above, is the gyromagnetic ratio. The value
$m_L\equiv1/(2\gamma^2M_0H_E)$ plays a role of the ``inertia
mass''; it depends upon the spin-flip field of the exchange
nature, $2H_E$, that characterises the intersublattice coupling.

In contrast to FMs, the nondissipative dynamics of AFMs can be
also described within the Hamiltonian formalism with the following
generalised energy (Hamilton function\footnote{~In order to
distinguish between the Hamilton function, $\mathcal{H}_{\mathrm
AFM}$, that generates equations of motion and the energy,
$E_{\mathrm AFM}$, as a dynamic variable (see below) we use
different notations.}) obtained from (\ref{Lagrangian_AFM}):
\begin{equation}\label{eq_energy_momenta}
    \mathcal{H}_{\mathrm AFM}=\frac{1}{2m_L}\mathbf{P}_L^2-\gamma\left[\mathbf{P}_L\cdot(\mathbf{L}\times{\mathbf{H}})\right]+w_{\mathrm{an}}(\mathbf{L}).
\end{equation}
Here the generalised momentum
$\mathbf{P}_L\equiv\partial\mathcal{L}_{\mathrm
AFM}/\partial\dot{\mathbf{L}}$ is canonically conjugated to the
generalised coordinate $\mathbf{L}$.

As can be directly seen from (\ref{eq_energy_momenta}), the
generalised energy of AFM particle, $\mathcal{H}_{\mathrm AFM}$,
includes both kinetic (first term) and potential (last term)
contributions. It means that the dynamics and kinetics of this
system is described within the \emph{phase} (\emph{vs}
configuration for FMs) space.

Dissipation is modeled with the
 Raileigh function which in the presence of spin-polarised current $J$ takes a form
\cite{gomo:2010}:
\begin{equation}
\label{Relay} \mathcal{R}_{\mathrm AFM} =
\gamma_{\mathrm{AFM}}m_L\dot{\mathbf{L}}^2-\frac{\sigma J}{2\gamma
M_0}\left[\mathbf{p}_{\mathrm
{curr}}\cdot(\mathbf{L}\times\dot{\mathbf{L}})\right].
\end{equation}
Here the first term models the internal damping, damping
coefficient $2\gamma_{\mathrm{AFM}}$ is the AFMR linewidth, the
constant $\sigma =\hbar
\gamma\varepsilon/(2eM_{0}v_{\mathrm{AFM}})$ is proportional to
the efficiency $\varepsilon$ of the spin transfer processes,
$v_{\mathrm{AFM}}$ is the volume of AFM nanoparticle, $\hbar$ is
the Plank constant, $e$ is the electron charge. Unit vector
$\mathbf{p}_{\mathrm {curr}}$ is parallel to the direction of the
current spin polarisation.

Thus, the stochastic equations for AFM in the phase space
$\{\mathbf{L},\mathbf{P}_L\}$ obtained from (\ref{Lagrangian_AFM})
and (\ref{Relay}) with substitution $\mathbf{H}\rightarrow
\mathbf{h}(t)$ acquire the form:
\begin{eqnarray}\label{eq_stochastic_Langevin}
  \dot{\mathbf{L}} &=& \mathbf{P}_L/m_L -\gamma\mathbf{L}\times\mathbf{h}\\
  \dot{\mathbf{P}}_L &=& \mathbf{F}_L+\mathbf{F}_{\mathrm{diss}}
  -\gamma\left(\mathbf{P}_L-2\gamma_{\mathrm{AFM}}m_L\mathbf{L}\right)\times\mathbf{h},\nonumber
\end{eqnarray}
where $\mathbf{F}_L\equiv-\partial
w_{\mathrm{an}}(\mathbf{L})/\partial \mathbf{L}$ is the potential
(gradient) force, and the dissipative force
$\mathbf{F}_{\mathrm{diss}}$ is given by the following expression
\begin{equation}\label{eq_dissipative_force}
    \mathbf{F}_{\mathrm{diss}}\equiv-\left.\frac{\partial \mathcal{R}_{\rm AFM}}{\partial\dot{\mathbf{L}}}\right|_{\dot{\mathbf{L}}\rightarrow \mathbf{P}_L}=-2\gamma_{\mathrm{AFM}}\mathbf{P}_L-\frac{\sigma
J}{2\gamma M_0}\mathbf{p}_{\mathrm {curr}}\times\mathbf{L}.
\end{equation}

Equations (\ref{eq_stochastic_Langevin}) describe the evolution of
AFM vector in the presence of thermal noise and in this sense are
analogous to the stochastic Eqs.~(\ref{eq-Landau_Lifshits}) for
magnetisation $\mathbf{M}$ of FMs. Both sets of equations (for
both FMs and AFMs) are linear in the random magnetic field
$\mathbf{h}$. For an AFM system this fact is nonobvious and can be
explained by the presence of small but nonzero macroscopic
magnetisation
$\mathbf{M}_{\mathrm{AFM}}\equiv\mathbf{M}_1+\mathbf{M}_2$ that in
the compensated AFM has a dynamic origin \cite{Andreev:1980} and
can be expressed in terms of the N\'eel vector:
$\mathbf{M}_{\mathrm{AFM}}\propto\dot{\mathbf{L}}
\times\mathbf{L}\propto\mathbf{P}_L \times\mathbf{L}$. On the
other hand, the noise terms in both (FMs and AFMs) equations are
\emph{multiplicative} and this can, in principle, result in a
possible stochastic resonance.

It should be stressed that though the AFM dynamics is similar to
the dynamics of point mass, Eqs.~(\ref{eq_stochastic_Langevin})
have one peculiarity compared with the standard Langevin equations
for a Brownian particle in a potential well. Namely, the first of
Eq.~(\ref{eq_stochastic_Langevin}) includes the noise and does not
include any dissipation term. This means that within the accepted
model of dissipation (and noise) there is no time-scale separation
between relaxation of generalised coordinate and generalised
momentum. This fact is a direct consequence of limitations on
$|\mathbf{L}|$ imposed by assumption of strong exchange coupling
and absence of the exchange relaxation. However, the
characteristic energy of exchange coupling is of the order of the
N\'eel temperature. So, for low (compared with the N\'eel)
temperatures and relatively small (compared with $H_E$) external
fields Eqs.~(\ref{eq_stochastic_Langevin}) give an adequate
description of AFM vector behaviour.

The Langevin Eqs.~(\ref{eq_stochastic_Langevin}) generate the
Fokker-Planck equation for AFM probability distribution function
$f(\mathbf{L},\mathbf{P}_L;t)$ in the phase space:
\begin{eqnarray}\label{eq_Fokker-Planck}
\frac{\partial f}{\partial
t}&=&\gamma^2\nabla_\mathbf{L}\cdot\left[\left(-2M_0H_E\mathbf{P}_L+D\hat\Lambda^\mathbf{L}\cdot\nabla_\mathbf{L}-\mathbf{P}_L\otimes\mathbf{L}\cdot\nabla_{\mathbf{P}_L}\right)f\right]\nonumber\\
&+&\nabla_{\mathbf{P}_L}\cdot\left[\left(\mathbf{F}_L+\mathbf{F}_{\mathrm{diss}}+D\gamma^2\hat\Lambda^{\mathbf{P}_L}\cdot\nabla_{\mathbf{P}_L}-\mathbf{L}\otimes\mathbf{P}_L\cdot\nabla_\mathbf{L}\right)f\right],
\end{eqnarray}
where we introduced the symbol $\hat\Lambda^{\mathbf{a}}\equiv\hat
1{\mathbf{a}}^2-{\mathbf{a}}\otimes{\mathbf{a}}$ with
${\mathbf{a}}=\mathbf{L}$ or $\mathbf{P}_L$, and omitted small
noise terms with $\gamma_{\mathrm{AFM}}$ for the sake of clarity.

Fokker-Planck Eq.~(\ref{eq_Fokker-Planck}) for AFM nanoparticles
is, in fact, the main result of this paper. It is much more
complicated than the analogous equations for FMs and (in contrast
to FMs) could not be solved in general case even for stationary
conditions. In the next section we consider some limiting cases
that allow to find approximate stationary solutions,
$f(\mathbf{L},\mathbf{P}_L)$, and evaluate $D$ from
fluctuation-dissipation theorem.
\section{Antiferromagnet probability distribution in the presence of spin-polarised current}\label{sec_Fokker-Plank}
The phase space of AFM particle is four-dimensional and this
substantially complicates analysis of Eq.~(\ref{eq_Fokker-Planck})
in general case. However, in some cases the effective
dimensionality can be reduced. The simplest case concerns the
system in the vicinity of equilibrium where all the possible
motions of AFM vector could be represented in terms of two
noninteracting normal modes with the amplitudes $c_\pm$ and
angular phases $\varphi_\pm$. If, in addition, we neglect
inhomogenuity in the phase $\varphi_\pm$ distribution, then,
distribution function can be factorized as:
$f(\mathbf{L},\mathbf{P}_L;t)=f_+(c_+;t)f_-(c_-;t)$. In what
follows we consider the case of AFM with the degenerate excitation
spectra for which two normal modes correspond to
clockwise/counter-clockwise rotations of AFM vector around $z$
axis with the frequency $\Omega_{\mathrm{AFMR}}$ (that is close to
AFMR frequency).

Spin current polarised along $z$ axis
($\mathbf{p}_{\mathrm{curr}}\|z$) interacts with both modes thus
enhancing the effective damping of one (say, ``+'') and
diminishing the effective damping of the other (say, ``--'')
\cite{gomo:2010}. In the subcritical regime
($|J|<J_{\mathrm{crit}}\equiv2\gamma_{\mathrm{AFM}}\Omega_{\mathrm{AFMR}}/(\gamma\sigma
H_E )$, positive damping) the static equilibrium state is stable
and normal modes are still well separated. In the supercritical
regime, $|J|>J_{\mathrm{crit}}$, an amplitude of one of the mode
growth to saturation value and the stable state corresponds to
rotation of AFM vector in $xy$ plane with the current-dependent
frequency $\omega=J\Omega_{\mathrm{AFMR}}/J_{\mathrm{crit}}$
\cite{gomo:2010}. Another normal mode corresponds to small
oscillations of AFM vector in $z$ direction, so, again, both modes
are well separated. Thus, the behaviour of AFM vector in the
subcritical and supercritical regions can be really described in
approximation of two independent normal modes.

To obtain the Fokker-Planck equations for $f(c_\pm)$ we use the
approach of energy representation for nonequilibrium Brownian-like
systems developed in \cite{Lev:2010PhRvE..82c1101L}. To this end
let us start from the Langevin equation for the energy $E_{\mathrm
AFM}\equiv\mathbf{P}_L^2/(2m_L)+w_{\mathrm{an}}(\mathbf{L})$
(compare with (\ref{eq_energy_momenta})):
\begin{equation}\label{eq_energy_Langevin}
  \frac{dE_{\mathrm
AFM}}{dt}=-\mathbf{P}_L\cdot\mathbf{F}_{\mathrm{diss}}+\gamma\left[\left(2\gamma_{\mathrm{AFM}}\mathbf{P}_L-\frac{\partial
w_{\mathrm{an}}}{\partial \mathbf{L}
}\right)\cdot\mathbf{L}\times\mathbf{h}\right].
\end{equation}
where the summands in the r.h.s. of Eq.~(\ref{eq_energy_Langevin})
should be expressed in terms of $E_{\mathrm AFM}$.

In approximation of noninteractive normal modes
Eq.~(\ref{eq_energy_Langevin}) is applicable to the energy $E_\pm=
4M^2_0\Omega_{\mathrm{AFMR}}^2m_Lc_\pm^2$ of \emph{each} mode.
Moreover, within the accepted approximation (fixed oscillation
frequency $\Omega_{\mathrm{AFMR}}$) $E_\pm$ could be considered as
the dynamic variables (that are proportional to the ``true''
canonical variables, actions).

 In
the subcritical region, $|J|<J_{\mathrm{crit}}$,
Eq.~(\ref{eq_energy_Langevin}) can be rewritten as follows:
\begin{eqnarray}\label{eq-Langevin_amp_subcrit}
  \frac{dc_{\pm}}{dt}&=&-\gamma_{\mathrm{AFM}}\left(1\pm\frac{J}{J_{\mathrm{crit}}}\right)c_{\pm}-\gamma\frac{\gamma_{\mathrm{AFM}}}{\Omega_{\mathrm{AFMR}}}c_{1,2}h_z\\
  &+&2\gamma\frac{\gamma_{\mathrm{AFM}}}{\Omega_{\mathrm{AFMR}}}\left(h_x\cos\Omega_{\mathrm{AFMR}}t+h_y\sin\Omega_{\mathrm{AFMR}}t\right)\nonumber\\
  &+&\gamma\left(h_x\sin\Omega_{\mathrm{AFMR}}t-h_y\cos\Omega_{\mathrm{AFMR}}t\right).\nonumber
\end{eqnarray}
It should be stressed that the same equation could be obtained
directly from (\ref{eq_stochastic_Langevin}) after transition to
amplitude-phase representation.

As it is seen from Eq.~(\ref{eq-Langevin_amp_subcrit}), the sign
$\pm$ corresponds to different modes which interact with the
current in different ways. If $J>0$, the effective damping of the
first mode (with the amplitude $c_+$) increases and that of the
second (with the amplitude $c_-$) decreases, due to the action of
spin-polarised current.

 Analysis of
Eq.~(\ref{eq-Langevin_amp_subcrit}) shows that one component
 of the random magnetic field $\mathbf{h}$, namely, that, which is
perpendicular to the plane of $\mathbf{L}$ rotation ($h_z$ in our
notations), is a source of multiplicative noise. However, if the
damping is rather small, $\gamma_{\mathrm{AFM}}\ll
\Omega_{\mathrm{AFMR}}$, the term with multiplicative noise can be
omitted. To this end Eq.~(\ref{eq-Langevin_amp_subcrit}) generates
the following Fokker-Planck equations:
\begin{equation}\label{eq_Fokker-PLank_modes}
  \frac{\partial f(c_\pm)}{\partial t}=\frac{\partial }{\partial
  c_\pm}\left[\gamma_{\mathrm{AFM}}\left(1\pm\frac{J}{J_{\mathrm{crit}}}\right)c_\pm f(c_\pm)+D\gamma^2\frac{\partial f(c_\pm)}{\partial
  c_\pm}\right].
\end{equation}
From the stationary solution of (\ref{eq_Fokker-PLank_modes}) one
gets the AFM probability distribution function $f(E_+,E_-)$:
\begin{equation}\label{eq_distribution_subcritical}
f(E_+,E_-)=f_0\exp{\left\{-\frac{\gamma_{\mathrm{AFM}}H_E}{D\Omega^2_{\mathrm{AFMR}}M_0}\left[\left(1+\frac{J}{J_{\mathrm{crit}}}\right)E_++\left(1-\frac{J}{J_{\mathrm{crit}}}\right)E_-\right]\right\}},
\end{equation}
where $f_{0}$ is a normalization constant.

In the absence of current the distribution
(\ref{eq_distribution_subcritical}) should coincide with the
Boltzmann distribution function. From the fluctuation-dissipation
theorem we get the diffusion coefficient for AFM particle
\begin{equation}\label{eq_diffus_constant_AFM}
  D_{\mathrm{AFM}}=\frac{\gamma_{\mathrm{AFM}}H_E}{\Omega^2_{\mathrm{AFMR}}M_0}T=\frac{1}{\gamma
  M_0}\frac{\gamma_{\mathrm{AFM}}}{\Omega_{\mathrm{AFMR}}}\sqrt{\frac{H_E}{H_a}}T,
\end{equation}
where $H_a\equiv\Omega^2_{\mathrm{AFMR}}/(\gamma^2 H_E)$ is the
field of magnetic anisotropy \cite{gomo:2010} which in the typical
AFMs is small compared with strong exchange field: $H_a\ll H_E$.

Remind, that the analogous coefficient for FM particle has a form
\cite{Garcia:PhysRevB.58.14937}:
\begin{equation}\label{eq_diffus_constant_FM}
  D_{\mathrm{FM}}=\frac{1}{\gamma
  M_0}\frac{\gamma_{\mathrm{FM}}}{\Omega_{\mathrm{FMR}}}T,
\end{equation}
where we used an explicit expression for the Gilbert damping
parameter through the frequency and half-width of FMR,
$\alpha_G\equiv\gamma_{\mathrm{FM}}/\Omega_{\mathrm{FMR}}$.
Comparing (\ref{eq_diffus_constant_AFM}) and
(\ref{eq_diffus_constant_FM}) one can easily see that the
diffusion coefficient in AFMs is greater that that for FMs due to
the large factor $\sqrt{H_E/H_a}\gg 1$, other things being equal.
This is one more manifestation of the above mentioned exchange
enhancement peculiar to AFM materials.

In the presence of spin-polarised current the distributions
(\ref{eq_distribution_subcritical}) are still Boltzmann-like with
two (instead of one for FMs) different effective temperatures for
each mode:
\begin{equation}\label{eq_effective temepratures}
  T_{\mathrm{eff}}^{\pm}=\frac{T}{1\pm J/J_{\mathrm{crit}}}.
\end{equation}
Expression (\ref{eq_effective temepratures}) shows that the
temperature of the ``soft'' mode (that one which becomes unstable
at $J\rightarrow J_{\mathrm{crit}}$) crucially growth, while the
temperature of the other mode diminishes. This fact illustrates
the current-induced energy swap between two modes. Seeming
singularity at $J\rightarrow  J_{\mathrm{crit}}$ is an artifact of
approximation which presupposes existence of high energy barrier
between the different stable states.

In the supercritical region one can get the distribution function
in a similar way. Neglecting, whenever it is possible, the small
value $\gamma_{\mathrm{AFM}}/\Omega_{\mathrm{AFMR}}\ll 1$, we
arrive at the following expression:
\begin{equation}\label{eq_distribution_supercritical}
f(E_+,E_-)=f_0\exp{\left[-\frac{4E_+}{[3+4(J/J_{\mathrm{crit}})^2]T}-\frac{\left(E_--E_{-}^{(0)}\right)^2}{2TE_{-}^{(0)}}\right]},
\end{equation}
where $E_+$ is related with oscillations of AFM vector in $z$
direction and the average energy of the second mode (related with
the rotation of AFM vector in $xy$ plane),
$E_{-}^{(0)}=M_0H_a(J/J_{\mathrm{crit}})^2$ is proportional to the
current value.

Like in FMs \cite{Slavin:2007ApPhL..91s2506T}, the distribution
(\ref{eq_distribution_supercritical}) is Gaussian-like with
respect to the energy of the second mode. However, in contrast to
FM, the half-width of corresponding  distribution is proportional
to $\Delta E_-\propto J$, so the ``quality factor''
$E_{-}^{(0)}/\Delta E_-\propto J$ growth with the current
value\footnote{~In the real system the current growth gives rise
to heating of the sample and thus imposes limitation on the
current value.}. Another peculiarity of AFM system compared with
FM is the presence of the additional energy fluctuations related
with the first mode.
\section{Conclusions}
In the present paper we have derived the Langevin and the
Fokker-Planck equations that take into account the peculiarities
of the dynamics of AFM (in contrast to magnetisation) vector.
These equations could be used for calculations of the dwell times
between the different states of AFM particle and the lineswidth of
resonances induced by external fields (including spin-polarised
current). It is shown that the thermal noise generated by
fluctuating magnetic field is multiplicative. As a result,
corresponding Fokker-Planck equation is nonlinear and this opens a
possibility for noise-induced transitions and stochastic
resonances in the system.

In the framework of the proposed approach we have calculated the
AFM energy  distribution function for the particular case of the
collinear two sublattice AFM in the presence of spin-polarised
current. We found that at a given temperature and quality factor
of the magnetic resonance the diffusion coefficient of AFM shows
an exchange enhancement compared to that of FM nanoparticle. It is
also shown that spin-polarised current affects the effective
temperatures of the normal oscillation modes in different ways: in
the subcritical region the temperature of the soft mode increases
and the temperature of the other (``hard'') mode decreases. In the
supercritical region the energy fluctuations of the soft mode grow
with respect to the current value slower than the average energy
of the mode. This opens a way to control the efficiency of energy
transfer from the current to AFM oscillator.

In our modeling we considered only the magnetic sources of noise.
However, in the presence of spin-polarised current the
fluctuations of the current value could be a source of
multiplicative noise, as seen e.g. from
Eq.~(\ref{eq-Langevin_amp_subcrit}). The problem of
current-induced noise needs a special treatment that accounts for
the relations between the magnetic state of AFM layer and
resistivity, Joule losses  etc.

Another important extension of the problems considered in the
present paper is seen in analysis of the possible current-induced
nonequilibrium states and their thermodynamics and information
characteristics in the spirit of recent general approaches
\cite{Horowitz:2010PhRvE..82f1120H,Hasegawa:2011PhRvE..84e1124H}.

\section{Acknowledgements}
The authors are grateful to Yu. B. Gaididei and B.I. Lev for
fruitful discussions. The paper was partially supported by the
grant from the Ministry of Education, Science, Youth and Sport of
Ukraine and by the Programme of Fundamental Researches of the
Department of Physics and Astronomy of National Academy of
Sciences of Ukraine.

\end{document}